\documentclass[twocolumn, superscriptaddress,showpacs,amsmath,amssymb,prl,aps,floatfix,longbibliography]{revtex4-1}
\usepackage[latin1]{inputenc}
\usepackage[american]{babel}
\usepackage[T1]{fontenc}
\usepackage{graphicx}
\usepackage{mathrsfs}
\usepackage{hyperref}
\usepackage{bm}
\usepackage{epstopdf}
\hypersetup{colorlinks=true, urlcolor=blue, citecolor=blue, filecolor=blue, linkcolor=blue}
\newcommand*{\diff}{\mathop{}\!\mathrm{d}}

\newcommand*{\Imm}{\mathop{}\!\mathbf{Im}}
\newcommand{\uimm}{\mathrm{i}}
\newcommand{\eu}{\mathrm{e}}
\newcommand{\daga}{^{\dagger}}

\begin{document}
\title{Phase reconstruction of strong-field excited systems \\by transient-absorption spectroscopy}
\author{Zuoye~Liu}
\email{liuzy04@yeah.net}
\affiliation{Max-Planck-Institut f\"ur Kernphysik, Saupfercheckweg 1, 69117 Heidelberg, Germany}
\affiliation{School of Nuclear Science and Technology, Lanzhou University, 730000, Lanzhou, China}
\author{Stefano~M.~Cavaletto}
\email{smcavaletto@gmail.com}
\affiliation{Max-Planck-Institut f\"ur Kernphysik, Saupfercheckweg 1, 69117 Heidelberg, Germany}
\author{Christian~Ott}
\affiliation{Max-Planck-Institut f\"ur Kernphysik, Saupfercheckweg 1, 69117 Heidelberg, Germany}
\author{Kristina~Meyer}
\affiliation{Max-Planck-Institut f\"ur Kernphysik, Saupfercheckweg 1, 69117 Heidelberg, Germany}
\author{Yonghao~Mi}
\affiliation{Max-Planck-Institut f\"ur Kernphysik, Saupfercheckweg 1, 69117 Heidelberg, Germany}
\author{Zolt\'an~Harman}
\affiliation{Max-Planck-Institut f\"ur Kernphysik, Saupfercheckweg 1, 69117 Heidelberg, Germany}
\author{Christoph~H.~Keitel}
\affiliation{Max-Planck-Institut f\"ur Kernphysik, Saupfercheckweg 1, 69117 Heidelberg, Germany}
\author{Thomas~Pfeifer}
\email{tpfeifer@mpi-hd.mpg.de}
\affiliation{Max-Planck-Institut f\"ur Kernphysik, Saupfercheckweg 1, 69117 Heidelberg, Germany}
\date{\today}
\begin{abstract}
We study the evolution of a V-type three-level system, whose two resonances are coherently excited and coupled by two ultrashort laser pump and probe pulses, separated by a varying time delay. We relate the quantum dynamics of the excited multi-level system to the absorption spectrum of the transmitted probe pulse. In particular, by analyzing the quantum evolution of the system, we interpret how atomic phases are differently encoded in the time-delay-dependent spectral absorption profiles when the pump pulse either precedes or follows the probe pulse. We experimentally apply this scheme to atomic Rb, whose fine-structure-split $5s\,^2S_{1/2}\rightarrow 5p\,^2P_{1/2}$ and $5s\,^2S_{1/2}\rightarrow 5p\,^2P_{3/2}$ transitions are driven by the combined action of a pump pulse of variable intensity and a delayed probe pulse. The provided understanding of the relationship between quantum phases and absorption spectra represents an important step towards full time-dependent phase reconstruction (quantum holography) of bound-state wave-packets in strong-field light-matter interactions with atoms, molecules and solids.
\begin{description}
\item[PACS numbers]
32.80.Qk, 32.80.Wr, 42.65.Re
\end{description}
\end{abstract}
\maketitle

A long-standing dream in the physics of light-matter interaction is the observation of the quantum dynamics of electrons bound in atoms and molecules \cite{RevModPhys.72.545, RevModPhys.81.163, nphoton.2014.48}. Traditional spectroscopy methods enable one to access the evolution of the amplitudes, related to the probability that a quantum system is in a given state. Much more demanding is the extraction of the associated quantum phases. 

\begin{figure}
\includegraphics[clip=true, width=\columnwidth]{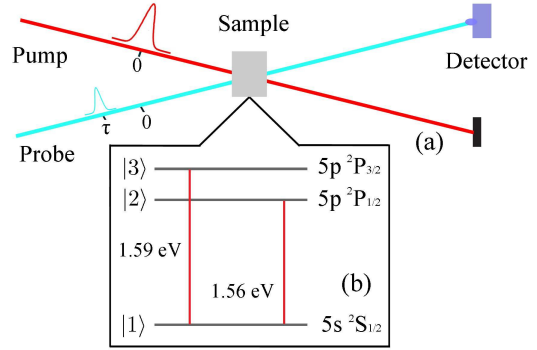}
\caption{(Color online) (a) Transient-absorption-spectroscopy setup in which, upon interaction of an atomic ensemble with two pulses separated by a time delay $\tau$, the absorption spectrum of the transmitted probe pulse is measured. (b) V-type three-level scheme used to model the ensemble of Rb atoms interacting with the laser pulses.}
\label{fig:Setup}
\end{figure}

With the development of femtosecond and attosecond laser pulses, transient-absorption spectroscopy (TAS) has emerged as a powerful method to gain amplitude and phase information on the evolution of a quantum system \cite{Mathies06051988, doi:10.1146/annurev.pc.43.100192.002433, PhysRevLett.98.143601, PhysRevLett.105.143002, GoulielmakisNature466, PhysRevLett.106.123601}. In such pump-probe geometry [Fig.~\ref{fig:Setup}(a)], a short pump pulse of tunable intensity excites a quantum system, whose ensuing dynamical evolution is measured by a weak probe pulse. The absorption spectrum of the transmitted field results from the interference of the incoming probe pulse with the electric field emitted by the system \cite{RevModPhys.40.441}. Since this interference pattern depends on the state of the system at the arrival of the probe pulse, its quantum-mechanical evolution can be time resolved by modifying the inter-pulse delay. %The absorption line at different time delays can thereby be linked to the dynamics of the quantum state, providing valuable information to reconstruct the evolution of the wave-packet. 
Furthermore, in contrast to most strong-field studies relying on the ionization of the system \cite{nature02277, PhysRevLett.105.053001}, TAS enables direct access to bound-state dynamics. 

Very recently, experiments started to investigate the absorption spectrum of a probe pulse which excites a coherent superposition of quantum states and is followed by a strong laser field \cite{PhysRevLett.105.143002, PhysRevA.86.063408,PhysRevLett.112.103001}. For nonautoionizing bound states, in the absence of this subsequent pulse, well known symmetric Lorentzian line shapes are observed. However, an additional strong pulse can be employed to nonlinearly drive the state generated by the probe field, thus manipulating the absorption lines into, e.g., asymmetric Fano-like shapes \cite{PhysRev.124.1866, Ott10052013, arxiv-1411.1545}. Although this method opens interesting prospects for line-shape quantum control \cite{nphoton.2014.113, 1367-2630-16-9-093005}, it still remains an open question how phase information is encoded in the time-delay-dependent strong-field absorption spectrum. %, as well as how similar probe-pump schemes can be used to complement the information obtained via traditional TAS studies.

Here, we investigate theoretically and experimentally the phase information which can be extracted from transient-absorption spectra for both above-mentioned scenarios, i.e., when the dynamics are either triggered or modified by an intense pump pulse. We use the V-type three-level scheme depicted in Fig.~\ref{fig:Setup}(b) %, whose two excited states represent an ideal number of levels for the investigation of fundamental strong-field-induced dynamics via time-delay-dependent absorption spectra. This scheme is employed 
to describe an ensemble of Rb atoms, with fine-structure-split excited states. From the ground state, the single-electron transitions $5s\,^2S_{1/2}\rightarrow 5p\,^2P_{1/2}$ ($|1\rangle\rightarrow|2\rangle$ at $794.76\,\mathrm{nm}$) and $5s\,^2S_{1/2}\rightarrow 5p\,^2P_{3/2}$ ($|1\rangle\rightarrow|3\rangle$ at $780.03\,\mathrm{nm}$) are excited by fs-duration pump/probe optical pulses centered at $780\,\mathrm{nm}$ and separated by a time delay $\tau$. We focus on the oscillations in $\tau$ exhibited by the absorption spectrum \cite{PhysRevA.65.043406} and show how these oscillating features differently encode the information on the intensity-dependent atomic phases for positive and negative time delays.

The atomic system is described by the density matrix $\hat{\rho}(t,\tau)$ \cite{Scully:QuantumOptics}, of elements $\rho_{ij}(t,\tau)$, $i,\,j\in\{1,\,2,\,3\}$. An off-diagonal element $\rho_{ij}(t,\tau)$, $i\neq j$, represents the coherence between states $|i\rangle$ and $|j\rangle$, while a diagonal element $\rho_{ii}(t,\tau)$ quantifies the population of level $|i\rangle$ at energy $\omega_i$, with transition energies $\omega_{ij} = \omega_i - \omega_j$ given by $\omega_{21} = 1.56\,\mathrm{eV}$ and $\omega_{31} = 1.59\,\mathrm{eV}$. Here, atomic units are used unless otherwise stated. The probe pulse %, i.e., the pulse whose absorption spectrum is detected, 
is centered at $t=\tau$ and is modeled by the classical field $\boldsymbol{\mathcal{E}}_{\mathrm{pr}}(t,\tau) = \mathcal{E}_{\mathrm{pr},0}\,f(t-\tau)\,\cos[\omega_{\mathrm{L}}(t-\tau) + \phi_{\mathrm{pr}}]\hat{\boldsymbol{e}}_z$, with the unit polarization vector $\hat{\boldsymbol{e}}_z$ along the $z$ direction, the laser frequency $\omega_{\mathrm{L}} = 1.59\,\mathrm{eV}$, the initial phase $\phi_{\mathrm{pr}}=0$, and the peak field strength $\mathcal{E}_{\mathrm{pr},0} = \sqrt{8\pi\alpha I_{\mathrm{pr}}}$ related to the peak intensity $I_{\mathrm{pr}}$ and the fine-structure constant $\alpha$. The envelope function is modeled by $f(t) = \cos^2{(\pi t/T)}\,R(t/T)$, where $R(x) = \theta(x+1/2)-\theta(x-1/2)$ is defined in terms of the Heaviside step function $\theta(x)$. Consequently, $f(t)$ is nonvanishing in an interval of duration $T = \pi T_{\mathrm{FWHM}}/[2\arccos{(\sqrt[4]{1/2})}]$, with $T_{\mathrm{FWHM}} = 30\,\mathrm{fs}$ being the full width at half maximum of $f^2(t)$. Similarly, the pump pulse is modeled by $\boldsymbol{\mathcal{E}}_{\mathrm{pu}}(t) = \mathcal{E}_{\mathrm{pu},0}\,f(t)\,\cos(\omega_{\mathrm{L}}t + \phi_{\mathrm{pu}})\hat{\boldsymbol{e}}_z$, with initial phase $\phi_{\mathrm{pu}}=0$, peak field strength $\mathcal{E}_{\mathrm{pu},0}$, and peak intensity $I_{\mathrm{pu}}$. The pump pulse is centered at $t=0$, such that it precedes the probe pulse for $\tau>0$ (pump-probe) and follows it for $\tau<0$ (probe-pump). Both linearly polarized pulses couple the ground state $|1\rangle \equiv 5s\,^2S_{1/2}$, with magnetic quantum number $M = \pm 1/2$, to the two closely lying states $|2\rangle \equiv 5p\,^2P_{1/2}$ and $|3\rangle \equiv 5p\,^2P_{3/2}$, also with $M = \pm 1/2$ \cite{Foot:AtomicPhysics}. For the pulse intensities used, we consider only electric-dipole-($E1$-)allowed transitions with $\Delta M = 0$, for which the dipole-moment matrix elements are $\boldsymbol{d}_{1k} = d_{1k}\,\hat{\boldsymbol{e}}_z$, $k\in\{2,\,3\}$, with $d_{13} = d_{12} \sqrt{2}$ \cite{PhysRevA.65.043406,johnson2007atomic}. This allows us to introduce the two complex, time-dependent Rabi frequencies $\varOmega_{\mathrm{R}k}(t,\tau) = d_{1k} [\mathcal{E}_{\mathrm{pr},0}\,f(t-\tau)\,\eu^{-\uimm\omega_{\mathrm{L}}\tau}\,\eu^{\uimm\phi_{\mathrm{pr}}} + \mathcal{E}_{\mathrm{pu},0}\,f(t)\,\eu^{\uimm\phi_{\mathrm{pu}}}]$, such that $\boldsymbol{d}_{1k}\cdot[\boldsymbol{\mathcal{E}}_{\mathrm{pu}}(t) + \boldsymbol{\mathcal{E}}_{\mathrm{pr}}(t,\tau)] = [\varOmega_{\mathrm{R}k}(t,\tau)\eu^{\uimm\omega_{\mathrm{L}}t} + \varOmega^*_{\mathrm{R}k}(t,\tau)\eu^{-\uimm\omega_{\mathrm{L}}t}]/2$. %{\color{red}Although the position of the atoms has not been explicitly included, position-dependent effects for the noncollinear geometry in Fig.~\ref{fig:Setup}(a) are effectively accounted for, as discussed in the following}. %However, since pump and probe fields propagate in nonidentical directions $\hat{\boldsymbol{e}}_{\mathrm{pu}}$ and $\hat{\boldsymbol{e}}_{\mathrm{pr}}$, respectively, a particle at position $\boldsymbol{r}'$ is excited with an effective delay $\tau'(\tau, \boldsymbol{r}') = \tau - (\hat{\boldsymbol{e}}_{\mathrm{pr}}- \hat{\boldsymbol{e}}_{\mathrm{pu}})\cdot \boldsymbol{r}'/c$, with $\tau$ being the time delay for an atom at position $\boldsymbol{r} = \boldsymbol{0}$.

The atomic system is initially in its ground state, i.e., $\rho_{ij,0} = \delta_{i1}\,\delta_{j1}$, with the Kronecker symbol $\delta_{ii'}$. The dynamical evolution of the system results from the Liouville--von~Neumann equation $\diff\hat{\rho}/\diff t = -\uimm[\hat{H}_0 + \hat{H}_{\mathrm{int}}(t,\tau),\,\hat{\rho}(t,\tau)] - \frac{1}{2}\{\hat{\varGamma},\hat{\rho}(t,\tau)\}$, with $[\hat{A},\hat{B}] = \hat{A}\hat{B} - \hat{B}\hat{A}$ and $\{\hat{A},\hat{B}\} = \hat{A}\hat{B} + \hat{B}\hat{A}$ \cite{Scully:QuantumOptics}. Here, $\hat{H}_0 = \sum_{i=1}^3\omega_i\hat{\sigma}_{ii}$ is the electronic-structure Hamiltonian, with ladder operators $\hat{\sigma}_{ij} = |i\rangle\langle j|$, while 
%\begin{equation}
$\hat{H}_{\mathrm{int}}(t,\tau) = -\frac{1}{2}\sum_{k=2}^3\varOmega_{\mathrm{R}k}(t,\tau)\,\hat{\sigma}_{1k}\,\eu^{\uimm\omega_{\mathrm{L}} t} + \mathrm{H.\,c.}$
% \label{eq:interaction-Hamiltonian}
% \end{equation}
is the $E1$-light-matter-interaction Hamiltonian in the rotating-wave approximation. The %$3\times 3$ 
diagonal matrix $\hat{\varGamma}$ has elements $\varGamma_{ij} = \delta_{ij}\gamma_j$, with $\gamma_1=0$ and $\gamma_2 = \gamma_3 = 1/(500\,\mathrm{fs})$. These decay rates are much larger than the spontaneous decay rates of the system and effectively account for broadening effects at work in the experiment.

The measured optical-density (OD) absorption spectrum is given by $\mathscr{S}(\omega, \tau) = -\log[S_{\mathrm{pr,out}}(\omega, \tau)/S_{\mathrm{pr,in}}(\omega)]$, where $S_{\mathrm{pr,in}}(\omega)$ %is the spectral energy of the incoming probe pulse, while $S_{\mathrm{pr,out}}(\omega, \tau)$ is that of the transmitted probe pulse, after interaction with a gas of Rb atoms (initially or subsequently) excited by a pump pulse centered on $t=0$. 
and $S_{\mathrm{pr,out}}(\omega, \tau)$ are the spectral intensities of the incoming and transmitted probe pulse, respectively. The absorption spectrum is related to the single-particle dipole response of the system, i.e., the evolution of the atomic coherences $\rho_{12}(t,\tau)$ and $\rho_{13}(t,\tau)$, via \cite{RevModPhys.40.441}
\begin{equation}
\mathscr{S}_{\mathrm{1}}(\omega,\tau) \propto -\Imm\Biggl[\sum_{k=2}^3\int_{-\infty}^{\infty}\omega_{k1}|\boldsymbol{d}_{1k}\,|\rho_{1k}(t,\tau)\,\eu^{-\uimm\omega (t-\tau)}\,\diff t\Biggr]\,,
\label{eq:S1}
\end{equation}
with Fourier transforms centered at the arrival time of the probe pulse. In the presence of a single pulse, the system, excited at $t=t_0$ into the state $\hat{\rho}(t_0)$, decays with time-dependent coherences $\rho_{1k}(t) = \rho_{1k}(t_0)\,\eu^{\uimm\omega_{k1}(t - t_0)}\,\eu^{-\frac{\gamma_k}{2}(t-t_0)}$, giving rise [Eq.~(\ref{eq:S1})] to stationary lines centered on the transition energies $\omega_{21}$ and $\omega_{31}$, with linewidths given by the decay rates $\gamma_{2}$ and $\gamma_{3}$, respectively. However, if two pulses are used, the coherences $\rho_{1k}(t,\tau)$ become explicitly time-delay-dependent, implying a modulation in $\tau$ of intensity and shape of the spectral lines. Although rapid oscillations in $\tau$ characterize the absorption spectrum in Eq.~(\ref{eq:S1}), these fast oscillating features are averaged out for the noncollinear experimental geometry in Fig.~\ref{fig:Setup}(a). %cannot be experimentally resolved in typical TAS experiments owing to the finite interaction volume. The experimental absorption spectrum at a given time delay $\tau$ is namely the average of contributions due to atoms at different positions $\boldsymbol{r}'$, associated with different effective time delays $\tau'(\tau, \boldsymbol{r}')$. Hence, a finite interaction volume $\Delta \boldsymbol{r}$ determines a finite time window $\Delta\tau$ over which time-delay-dependent features can be experimentally distinguished. 
We include this effect in our model via convolution of the spectrum in Eq.~(\ref{eq:S1}) with a normalized Gaussian function $G(\tau, \Delta\tau)$ of width $\Delta\tau = 5\times2\pi/\omega_{\mathrm{L}}$, thus obtaining
\begin{equation}
\mathscr{S}(\omega,\tau) = \langle \mathscr{S}_{\mathrm{1}}(\omega,\tau)\rangle_{\tau} = \int_{-\infty}^{\infty}G(\tau - \tau',\Delta\tau)\,\mathscr{S}_{\mathrm{1}}(\omega,\tau')\,\diff \tau' .
\label{eq:spectrum}
\end{equation}

\begin{figure}
\includegraphics[clip=true, width=\columnwidth]{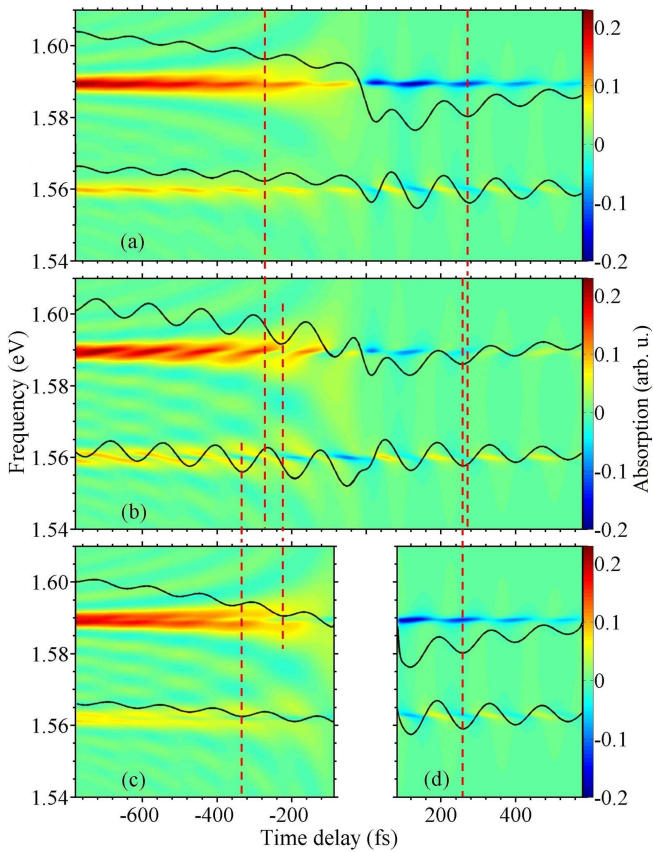}
% \centering\input{Images/source.tex}
% \includegraphics[clip=true, width=\columnwidth]{myFig2.eps}
\caption{(Color online) Theoretical absorption spectra of a delayed probe pulse of intensity $9.2\times 10^{8}\,\mathrm{W/cm^2}$ interacting with an ensemble of Rb atoms, modeled via the three-level system in Fig.~\ref{fig:Setup}(b) and driven at $t =0$ by a pump pulse of intensity (a) $1.0\times 10^{10} \,\mathrm{W/cm^2}$ and (b) $2.8\times 10^{10} \,\mathrm{W/cm^2}$. Panels~(c) and (d) display, for negative and positive time delays, respectively, absorption spectra for a pump pulse of $1.0\times 10^{10} \,\mathrm{W/cm^2}$ and artificial phases $\phi_2$ and $\phi_3$ [Eq.~(\ref{eq:artificial})], with (c) $\phi_2 = \phi_3 =-2.26$ for $\tau<0$, and (d) $\phi_2 = 0$, $\phi_3 = 0.67$ for $\tau>0$. In each panel, the top (bottom) black lines represent the absorption spectra evaluated at the transition energy $\omega_{31}$ ($\omega_{21}$) in arbitrary units. All black lines are on the same scale, with the 0 aligned on the corresponding transition energy.}
\label{fig:theory}
\end{figure}

In order to investigate the intensity-dependent phase information which can be extracted from $\mathscr{S}(\omega,\tau)$, in Fig.~\ref{fig:theory} we compare numerical calculations of Eq.~(\ref{eq:spectrum}) for a probe intensity of $9.2\times 10^{8}\,\mathrm{W/cm^2}$ and different pump intensities. As shown in Fig.~\ref{fig:theory}(a) for a peak intensity of $1.0\times 10^{10} \,\mathrm{W/cm^2}$, the spectrum consists of two lines centered on the $E1$-allowed-transition energies $\omega_{21}$ and $\omega_{31}$. %Only for negative time delays the absorption spectrum at fixed $\tau$ displays additional maxima and minima as a function of $\omega$ \cite{0953-4075-47-12-124008}, whose number, as first studied in semiconducting samples~\cite{PhysRevLett.59.2588}, increases with rising values of $|\tau|$. 
Furthermore, for both positive and negative time delays, amplitude and shape of the two absorption lines oscillate in $|\tau|$, with a periodicity determined by the beating frequency $\omega_{32} = 2\pi/(140\,\mathrm{fs})$ and a discontinuity across $\tau =0$, where the pulses overlap. %This behavior can be explained by considering that $\rho_{12}$ and $\rho_{13}$ oscillate at the respective transition energies $\omega_{21}$ and $\omega_{31}$, such that the periodicity with which the second arriving pulse---pump pulse for $\tau<0$, probe pulse for $\tau>0$---encounters these two fast oscillating functions with an identical phase relation is given by the beating frequency $\omega_{32}$. 
The black lines in Fig.~\ref{fig:theory}(a) highlight this behavior, by displaying $\mathscr{S}(\omega_{21},\tau)$ and $\mathscr{S}(\omega_{31},\tau)$ evaluated at the atomic transition energies. Absorption strengths are shown in arbitrary units, as they depend on the gas pressure in the cell, which is not included in the simulation.

Figure~\ref{fig:theory}(b) displays analogous features for a pump intensity of $2.8\times 10^{10} \,\mathrm{W/cm^2}$. Here, however, different phases of the time-delay-dependent oscillations can be recognized. This is apparent from the time delays at which $\mathscr{S}(\omega_{21},\tau)$ and $\mathscr{S}(\omega_{31},\tau)$ exhibit, e.g., their minima, as highlighted by the red, dashed lines. For $\tau>0$, these minima are aligned both in Fig.~\ref{fig:theory}(a) and \ref{fig:theory}(b), although a common phase shift can be distinguished. In contrast, for $\tau<0$, the minima are aligned in Fig.~\ref{fig:theory}(a), while in Fig.~\ref{fig:theory}(b) they appear shifted towards opposite directions as a consequence of the larger pump intensity. The different behavior displayed at positive and negative delays reflects the different excitation sequence undergone by the system. While at $\tau>0$ the pump pulse generates the initial coherences, for $\tau<0$ it strongly modifies the coherent superposition of states generated by the probe pulse.

In Figs.~\ref{fig:theory}(c) and \ref{fig:theory}(d) we show that the phase of the oscillations in $\tau$ encodes the key information about the intensity-dependent phases of the atomic system. In both panels, a pump intensity of $1.0\times 10^{10} \,\mathrm{W/cm^2}$ is used as in Fig.~\ref{fig:theory}(a). Here, however, we modify the phases of the atomic system by artificially transforming the density matrix $\hat{\rho}(T/2,\tau)$---describing the state of the system at $t = T/2$ immediately after the interaction with the pump pulse---into
\begin{equation}
\hat{\rho}'(T/2,\tau) = \hat{\varPhi}\,\hat{\rho}(T/2,\tau)\,\hat{\varPhi}\daga.
\label{eq:artificial}
\end{equation}
The usual equations of motion are employed to calculate the subsequent dynamics of the system. In Eq.~(\ref{eq:artificial}), $\hat{\varPhi}$ is a diagonal matrix of elements $\varPhi_{ij} = \eu^{-\uimm\phi_i}\,\delta_{ij}$, with $\phi_1 = 0$, while $\phi_2$ and $\phi_3$ represent shifts in the phase of levels $|2\rangle$ and $|3\rangle$, respectively, such that $\rho_{ij}'(T/2,\tau) = \rho_{ij}(T/2,\tau)\,\eu^{-\uimm(\phi_i-\phi_j)}$. 

These artificial shifts $\phi_k$ effectively model phase changes undergone by the system at varying pump intensities, to which the time-delay-dependent absorption spectra are sensitive. For instance, in Fig.~\ref{fig:theory}(c), each $\phi_k$ is set equal to the difference between the phase changes caused by a pump intensity of $2.8\times 10^{10} \,\mathrm{W/cm^2}$ [Fig.~\ref{fig:theory}(b)] and of $1.0\times 10^{10} \,\mathrm{W/cm^2}$ [Fig.~\ref{fig:theory}(a)], which can be extracted from the corresponding dynamics of the density matrix. The resulting absorption spectrum features time-delay-dependent oscillations with the same amplitudes as in Fig.~\ref{fig:theory}(a), but whose phases are now equal to those displayed in Fig.~\ref{fig:theory}(b). An analogous procedure is applied in Fig.~\ref{fig:theory}(d) for positive time delays, with new phases $\phi_k$.

\begin{figure}
\includegraphics[clip=true, width=\columnwidth]{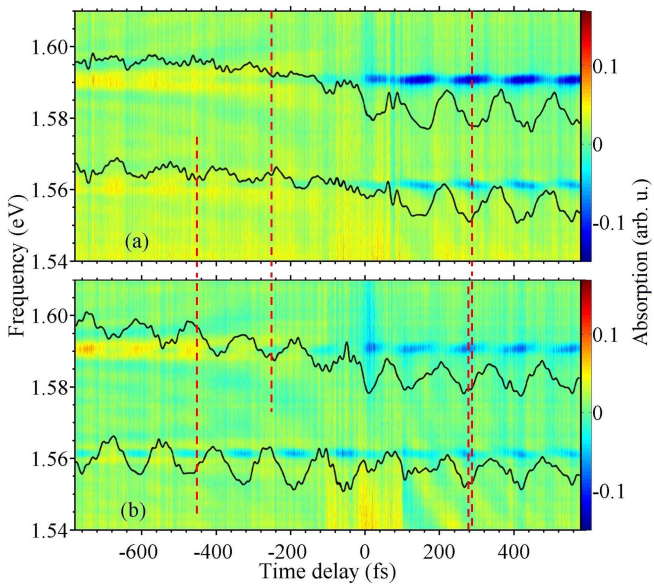}
\caption{(Color online) Experimental absorption spectra of a delayed probe pulse of intensity $I_{\mathrm{pr}} = 9.2\,\times\,10^8\,\mathrm{W/cm^2}$, transmitted through an ensemble of Rb atoms driven at $t =0$ by a pump pulse of intensity (a) $I_{\mathrm{pu}} = 1.0\,\times\,10^{10}\,\mathrm{W/cm^2}$ and (b) $I_{\mathrm{pu}} = 2.8\,\times\,10^{10}\,\mathrm{W/cm^2}$. The black on-resonance lineouts have the same meaning as in Fig.~\ref{fig:theory}.}
\label{fig:experiment}
\end{figure}

In order to test the relation between intensity-dependent atomic phases and time-delay-dependent absorption spectra, in Fig.~\ref{fig:experiment} we show results from the experimental measurement of the OD absorption spectrum of a probe pulse transmitted through an ensemble of Rb atoms. A Ti:Sapphire laser is utilized, with a repetition rate of $4\,\mathrm{kHz}$, pulse energy of $0.7\,\mathrm{mJ}$, FWHM of $30\,\mathrm{fs}$, and central wavelength of $780\,\mathrm{nm}$. The laser beam is split by a spatial mask with two irises into two beams, which are used as probe and pump beams in Fig.~\ref{fig:Setup}(a). A delay line is included into the pump beam path by means of a piezo translation stage at close-to-normal incidence, allowing one to scan the time delay between the two pulses. Rb vapor is supplied in a cell with $20\,\mathrm{mm}$ length, which is heated to $160\,\mathrm{^{o}C}$ by a home-made heating and control system. Using the known vapor pressure of Rb, the atomic density is estimated to be $2.8\times10^{14}\,\mathrm{cm}^{-3}$. The pump and probe pulses, with beam sizes of $3.5\,\mathrm{mm}$ and $3.0\,\mathrm{mm}$, respectively, are focused by a concave mirror with a focusing length of $500\,\mathrm{mm}$. After passing through the sample, the probe beam is picked up by an identical concave mirror and coupled into a fiber-pigtailed spectrometer (Ocean Optics, USB2000+). The probe pulse has an intensity of $9.2\,\times\,10^8\,\mathrm{W/cm^2}$, and Figs.~\ref{fig:experiment}(a) and \ref{fig:experiment}(b) show experimental results for a pump intensity of $1.0\,\times\,10^{10}\,\mathrm{W/cm^2}$ and $2.8\,\times\,10^{10}\,\mathrm{W/cm^2}$, respectively. The absorption lines at each resonance are broadened to a measurable width as a result of Doppler and collision-induced pressure broadening, as well as additional broadening in the nanosecond pedestal of the femtosecond pulses. For low pump intensities [Fig.~\ref{fig:experiment}(a)], at $\tau<0$, the small amplitude of the oscillations in $\tau$ renders the associated phases not straightforwardly extractable. Apart from that, the experimental results are in good agreement with the theoretical predictions from Fig.~\ref{fig:theory}, showing analogous intensity-dependent phase shifts in the oscillating features of the absorption spectra.

\begin{figure}
\includegraphics[width=\columnwidth, keepaspectratio]{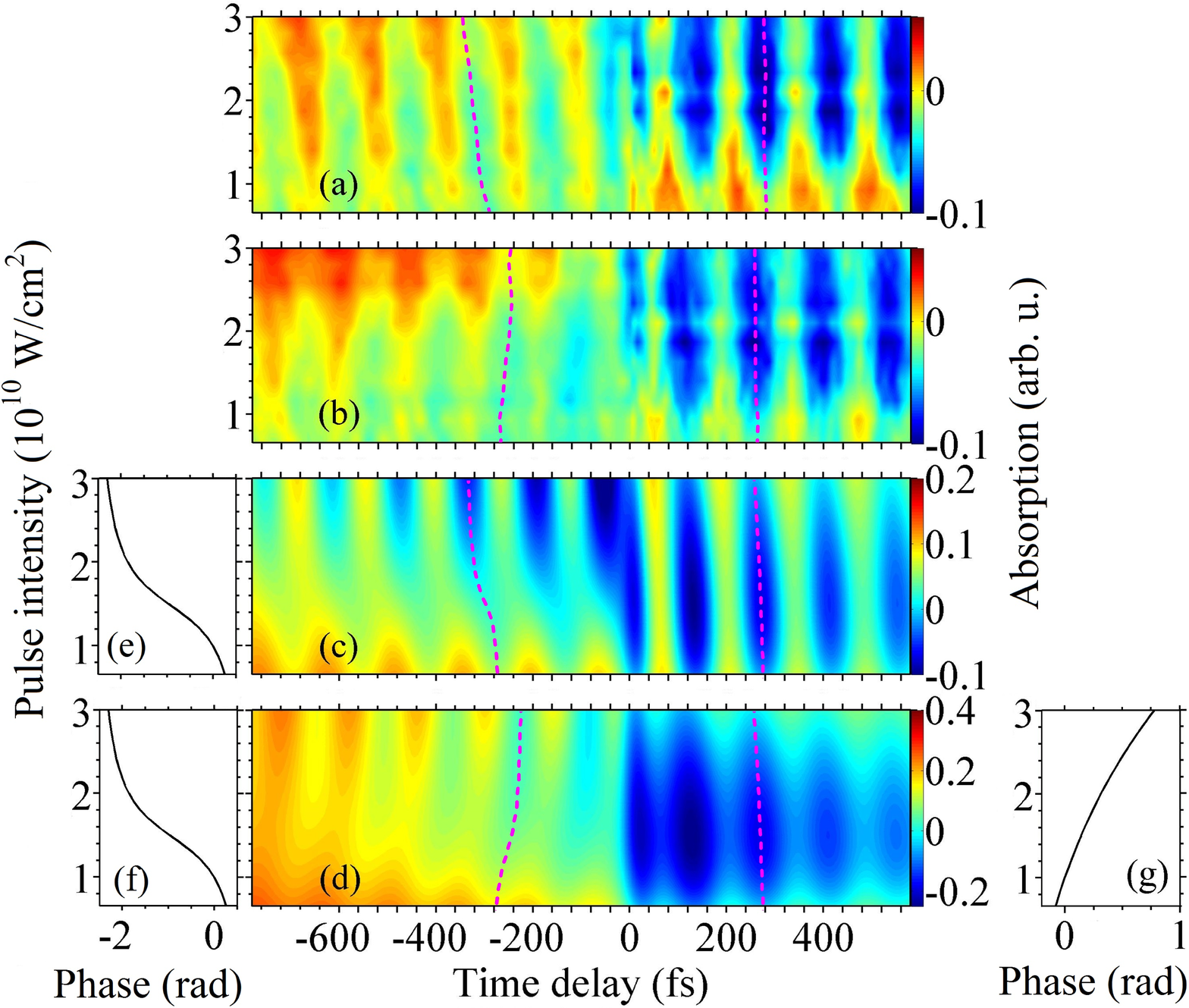}
\caption{(Color online) [(a)-(b)] Experimental and [(c)-(d)] theoretical absorption spectra at fixed frequencies [$\omega_{21} = 1.56\,\mathrm{eV}$ in (a) and (c), $\omega_{31} = 1.59\,\mathrm{eV}$ in (b) and (d)] of a delayed probe pulse of intensity $I_{\mathrm{pr}} = 9.2\,\times\,10^8\,\mathrm{W/cm^2}$, transmitted through an ensemble of Rb atoms driven at $t =0$ by a pump pulse of varying intensity. The dashed, magenta lines display the time delays (as a function of pump intensity) associated with minima in the absorption spectra. Panels~(e), (f) and (g) show the artificial phases [$\phi_2$ in (e), $\phi_3$ in (f) and (g), see also Eq.~(\ref{eq:artificial})] which need be added at $I_{\mathrm{pu}} = 1.0\times 10^{10}\,\mathrm{W/cm^2}$ to reproduce the intensity-dependent phases of the time-delay-dependent oscillations at [(e)-(f)] negative and (g) positive time delays.}
\label{fig:intensity}
\end{figure}

Finally, we focus on the dependence of phase effects upon the intensity of the pump pulse. Figures~\ref{fig:intensity}(a) and \ref{fig:intensity}(b) display the experimental absorption spectra $\mathscr{S}(\omega_{21},\tau)$ and $\mathscr{S}(\omega_{31},\tau)$ at the transition energies $\omega_{21}$ and $\omega_{31}$, respectively, for varying pump intensities. %The periodic behavior exhibited as a function of time delay is subject to a clear intensity-dependent shift. 
These measurement results quantify intensity-dependent shifts in the phase of the time-delay-dependent spectra, which are in good agreement with the theoretical predictions in Figs.~\ref{fig:intensity}(c) and \ref{fig:intensity}(d) and confirm the observations from Figs.~\ref{fig:theory} and \ref{fig:experiment}. %For $\tau>0$, the minima of $\mathscr{S}(\omega_{21},\tau)$ and $\mathscr{S}(\omega_{31},\tau)$ remain aligned while shifting towards lower values of $|\tau|$ at increasing pump intensities. For $\tau<0$, however, these shifts point towards opposite directions. 

The experimentally observable phase properties displayed by $\mathscr{S}(\omega_{21},\tau)$ and $\mathscr{S}(\omega_{31},\tau)$ reflect the different physical mechanisms acting at negative and positive time delays. This is further analyzed %reflect qualitatively different information on the phase evolution of the quantum system for a pump-probe or probe-pump scheme. This is shown 
by implementing the artificial-phase method introduced in Eq.~(\ref{eq:artificial}) to reproduce intensity-dependent phase effects. For $\tau<0$, the pump pulse strongly modifies the excited state generated at $t=\tau$ by the probe pulse. The intensity-dependent phases of the oscillations in $\tau$ featured in Figs.~\ref{fig:intensity}(c) and \ref{fig:intensity}(d) enable one to quantify the shifts in the phase of the atomic coherences ensuing from the interaction with strong pump pulses of varying intensity, as captured by $\phi_2(I_{\mathrm{pu}})$ and $\phi_3(I_{\mathrm{pu}})$ in Figs.~\ref{fig:intensity}(e) and ~\ref{fig:intensity}(f), respectively. In contrast, for $\tau>0$, the pump pulse is responsible for the initial excitation of the system. The parallel shift undergone by $\mathscr{S}(\omega_{21},\tau)$ and $\mathscr{S}(\omega_{31},\tau)$ at increasing pump intensities [Figs.~\ref{fig:intensity}(c) and \ref{fig:intensity}(d)] is therefore a consequence of the dependence of the initial coherences---and their relative phase---upon the intensity of the pump pulse. Artificial phases can be used to effectively reproduce this behavior. In this case, however, the phases of the resulting spectra $\mathscr{S}(\omega_{21},\tau)$ and $\mathscr{S}(\omega_{31},\tau)$ are identically affected only by the phase difference $\phi_3(I_{\mathrm{pu}}) - \phi_2(I_{\mathrm{pu}})$, which is shown in Fig.~\ref{fig:intensity}(g) for $\phi_2(I_{\mathrm{pu}}) = 0$.

In conclusion, we have used a V-type three-level system to investigate strong-field-induced phase effects in transient-absorption spectra for both positive and negative time delays and for pump pulses of variable intensity. Furthermore, we have developed an artificial-phase method to interpret time-delay-dependent absorption spectra and relate them to the phase evolution and intensity dependence of atomic quantum coherences. %Different information can be accessed at positive and negative time delays. 
%In particular, at low intensities we showed that only at positive time delays, i.e., when the phase shift is added upon interaction with the probe pulse, this significantly affects the line shape. At increasing pump intensities, the absorption spectrum periodically oscillates as a function of the time delay, displaying symmetric and asymmetric line shapes. The phase of these oscillating features is also determined by the intensity of the pump pulse, with a different trend at positive and negative time delays. 
Although this method was here introduced to analyze intensity-dependent phase effects, the level-specific phases $\phi_k$ could also be employed to model additional interactions of the atomic system with external fields selectively modifying the two excited states. Understanding the phase-sensitive link between quantum dynamics and transient-absorption spectra represents an essential step towards the application of TAS to the extraction of quantum phases in bound-state atomic systems interacting with strong-field pulses and to sensitive tests of quantum-dynamics theory.

\begin{acknowledgments}
The authors acknowledge support from the Max-Planck Research Group program and the DFG (grant PF790/1-1). Z.~Liu acknowledges the scholarship award for excellent doctoral students granted by the Ministry of Education of China.
\end{acknowledgments}

\end{document}